\documentclass[aps,floatfix,showpacs,tightenlines,superscriptaddress,amsmath,amssymb]{revtex4}
\usepackage{amssymb,amsbsy,epsfig,color,graphicx}

\begin{document}

\title{Continuous variable quantum teleportation with sculptured and noisy non-Gaussian resources}

\author{F. Dell'Anno}
\affiliation{Dipartimento di Matematica e Informatica, Universit\`a
degli Studi di Salerno, Via Ponte don Melillo, I-84084 Fisciano (SA), Italy}
\affiliation{CNR-INFM Coherentia, Napoli, Italy, CNISM Unit\`a di Salerno,
and INFN Sezione di Napoli, Gruppo collegato di Salerno, Baronissi (SA), Italy}

\author{S. De Siena}
\affiliation{Dipartimento di Matematica e Informatica, Universit\`a
degli Studi di Salerno, Via Ponte don Melillo, I-84084 Fisciano (SA), Italy}
\affiliation{CNR-INFM Coherentia, Napoli, Italy, CNISM Unit\`a di Salerno,
and INFN Sezione di Napoli, Gruppo collegato di Salerno, Baronissi (SA), Italy}

\author{L. Albano Farias}
\affiliation{Dipartimento di Fisica, Universit\`a degli Studi di Salerno,
Via S. Allende, I-84081 Baronissi (SA), Italy}

\author{F. Illuminati}
\thanks{Corresponding author. Electronic address: illuminati@sa.infn.it}
\affiliation{Dipartimento di Matematica e Informatica, Universit\`a
degli Studi di Salerno, Via Ponte don Melillo, I-84084 Fisciano (SA), Italy}
\affiliation{CNR-INFM Coherentia, Napoli, Italy, CNISM Unit\`a di Salerno,
and INFN Sezione di Napoli, Gruppo collegato di Salerno, Baronissi (SA), Italy}
\affiliation{ISI Foundation for Scientific Interchange,
Viale Settimio Severo 65, 00173 Torino, Italy}

\date{February 19, 2008}

\begin{abstract}
We investigate continuous variable (CV) quantum teleportation using
relevant classes of non-Gaussian states of the radiation field as entangled resources.
First, we introduce the class two-mode squeezed symmetric superposition of Fock states,
including finite truncations of twin-beam Gaussian states as special realizations.
These states depend on a set of free independent parameters
that can be adjusted for the optimization of teleportation protocols,
with an enhancement of the success probability of teleportation
both for coherent and Fock input states.
We show that the optimization procedure reduces the entangled resources
to truncated twin beam states, which thus represents an optimal class
of non-Gaussian resources for quantum teleportation.
We then introduce a further class of two-mode non-Gaussian entangled resources,
in the form of squeezed cat-like states. We analyze the performance and the
properties of such states when optimized for (CV) teleportation,
and compare them to the optimized squeezed Bell-like states introduced
in a previous work \cite{CVTelepNoi}. We discuss how optimal resources for
teleportation are characterized by a suitable balance of entanglement content
and squeezed vacuum affinity.
We finally investigate the effects of thermal noise on the efficiency of
quantum teleportation. To this aim, a convenient framework is to describe
noisy entangled resources as linear superpositions of non-Gaussian state
and thermal states. Although the presence of the thermal component strongly
reduces the teleportation fidelity, noisy non-Gaussian states remain preferred
resources when compared to noisy twin-beam Gaussian states.
\end{abstract}

\maketitle

In recent years, much effort has been dedicated to the theoretical investigation and
experimental production of highly nonclassical, non-Gaussian states of the radiation field
\cite{PhysRep}. Such states are endowed with enhanced nonclassical properties,
such as entanglement and negative quasi-probability phase-space distribution,
that can be useful for improving the efficiency of quantum information and communication protocols \cite{PhysRep,KimBS,KitagawaPhotsub,DodonovDisplnumb,VanLoock,CxKerrKorolkova}.
Importantly, a general, rigorous result states that, at fixed covariance matrix,
some of these desirable properties are minimized by Gaussian states, which can then
be considered as extremal "benchmark" or lowest-threshold states for quantum
information applications \cite{ExtremalGaussian}.
Therefore, at least in principle, the exploitation of non-Gaussian states
and non-Gaussianity allows to quantitatively improve the performance
of several quantum information protocols.
For instance, non-Gaussian cloners of coherent states turn out to be optimal with respect
to Gaussian ones, exhibiting an enhanced single-clone fidelity \cite{Cerf}.
Concerning quantum teleportation with continuous variables (CV), the success
probability of CV teleportation can be greatly increased by using entangled non-Gaussian
resources, ''degaussified'' through conditional measurements, consisting of photon-subtraction  \cite{KitagawaPhotsub,Opatrny,Cochrane,Olivares,CVTelepNoi}.
In particular, as shown in Ref.~\cite{CVTelepNoi}, the implementation of simultaneous phase matched multiphoton processes and conditional measurements can be used to introduce a general class of two-mode non-Gaussian entangled states, in the form of squeezed Bell-like states
depending on an extra free, experimentally adjustable parameter.
The optimization on this additional free parameter allows a remarkable increase of the teleportation
fidelity for several relevant input states, compared both to Gaussian and other degaussified
resources \cite{CVTelepNoi}. From the point of view of quantum state engineering, several schemes
for the generation of non-Gaussian states have been proposed \cite{CxKerrKorolkova,AgarTara,DeGauss1,DeGauss2,DeGauss3,DeGauss4},
and some of them, based on photon addition or subtraction, have been experimentally
implemented to obtain photon-added and photon-subtracted single-mode states
from Gaussian inputs \cite{ZavattaScience,ExpdeGauss1,ExpdeGauss2}.
More recently, very advanced experimental realizations have been reported \cite{Grangier,BelliniProbing,GrangierCats}. First, enhancement of entanglement
and production of states with negative two-mode Wigner function have been
successfully obtained via photon subtraction of a single
delocalized photon from an initial two-mode Gaussian state \cite{Grangier}.
Remarkably, the photon-addition/subtraction operations,
performed on a thermal light field, have also led to the demonstration
of the commutation relation rules, one of the constitutive relations
of quantum mechanics \cite{BelliniProbing}.
Moreover, using homodyne detection and Fock-state resources,
the experimental generation of large squeezed Schr\"{o}dinger cat states
has been achieved \cite{GrangierCats}.
These optical cat-like states are of particular importance because
they are strongly resilient against decoherence \cite{SerafiniCats}. \\
In the present work, we aim to extend the results obtained in Ref.~\cite{CVTelepNoi}
by generalizing the class of squeezed Bell-like states and by proposing an alternative class
of non-Gaussian superposition states, and by showing that these two classes of
entangled non-Gaussian resources allow for the optimization of CV quantum teleportation.
Specifically, we investigate CV teleportation with two different classes
of two-mode entangled resources: Squeezed symmetric superpositions of Fock states,
and squeezed cat-like states.
These two classes share two common features: Both are associated to
the squeezing operator as CV generator of quantum correlations, and
for a trivial choice of the parameters, both reduce to the Gaussian twin beam (TwB).
The last part of the work will be dedicated to the investigation of the effects
of the presence of thermal noise on the performance of two-mode non-Gaussian states
used as resources for quantum teleportation.
In particular, we will consider the non-Gaussian resources obtained by superimposing
squeezed Bell-like states or squeezed cat-like states and two-mode thermal states.
Non-Gaussian states and thermal states are assumed to be independently generated,
and overlap in a common spatial volume \cite{Glauber}.
Due to the thermal contribution, the ensuing mixed states and their correlation
properties are modified by the presence of thermal photons.
We limit the discussion to the situation of ideal teleportation protocol, i.e. ideal
Bell measurements and decoherence-free propagation in communication channels.
Detailed analysis in the instance of the most general realistic situation,
including various sources of noise, will be discussed elsewhere. \\
The paper is organized as follows.
In Section \ref{SecSqueezTruncTwB}, we briefly recall the mathematical machinery
needed for the description of the ideal Braunstein-Kimble CV teleportation protocol
in the characteristic function formalism.
Next, we introduce the class of squeezed symmetric superposition of Fock states,
and apply to them the optimization procedure defined in Ref.~\cite{CVTelepNoi}.
In Section \ref{SecSquezCatlike}, we introduce the class od squeezed cat-like states,
we apply the optimization procedure, and we compare results with those obtained
in the previous Section.
In Section \ref{SecNonGaussMixed} we will analyze the effect
of noise on the performance of non-Gaussian resources for CV teleportation.
Finally, in Section \ref{Conclusions} we will draw our conclusions and discuss
some perspectives for future research.

\section{CV teleportation with squeezed symmetric superposition of Fock states}
\label{SecSqueezTruncTwB}

\begin{figure}[h]
\centering
\includegraphics*[width=17cm]{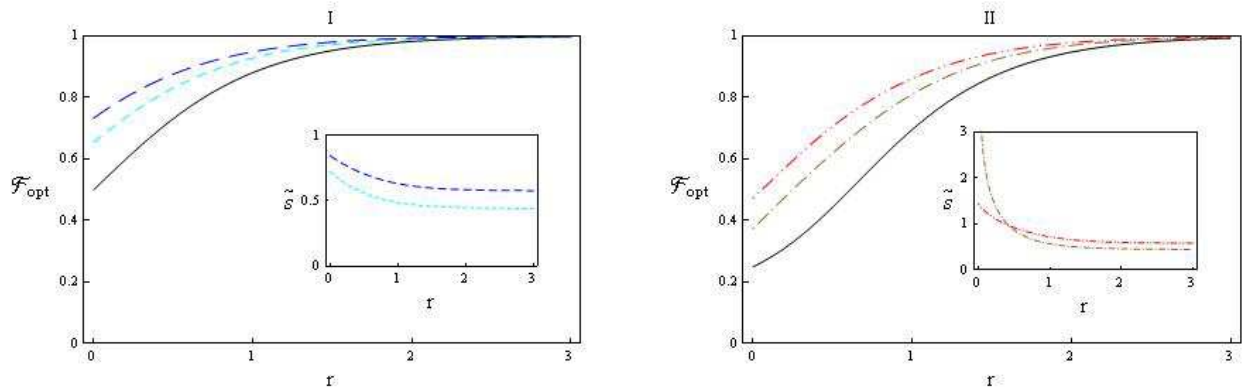}
\caption{Optimal fidelity of teleportation
$\mathcal{F}_{opt}$, as a function of the squeezing parameter
$r$, with $\phi=\pi$. In Panel I, we report the fidelity for the teleportation
of input coherent states $|\beta\rangle$
using different two-mode entangled resources:
$(a)$ Gaussian squeezed state resource (full line);
$(b)$ squeezed Bell-like state (dashed line);
$(c)$ squeezed symmetric superposition of Fock states (long-dashed line).
In Panel II, we report the fidelity for the teleportation
of input single-photon Fock states $|1\rangle$
using different two-mode entangled resources:
$(a)$ Gaussian squeezed state resource (full line);
$(b)$ squeezed Bell-like state (dot-dashed line);
$(c)$ squeezed symmetric superposition of Fock states (double-dot-dashed line).
In plot I the value of $\beta$ is arbitrary.
The insets in both Panels give the maximal values of the parameter $s=\tilde{s}$
as a function of $r$ for given entangled resource and fixed input state.
The plotstyles are chosen as specified above.}
\label{FigFidSculpTelep}
\end{figure}

\begin{figure}[h]
\centering
\includegraphics*[width=8cm]{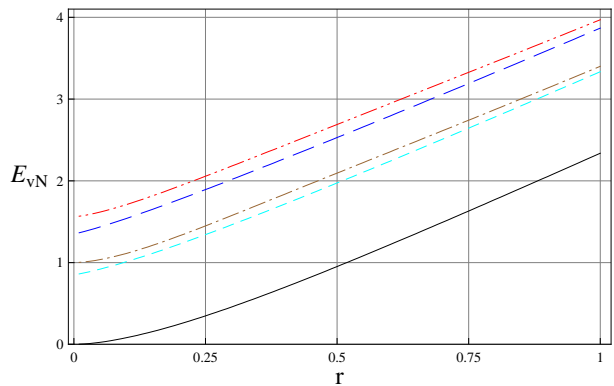}
\caption{Plot of the von Neumann entropy $E_{vN}$ as a function of $r$, with $s=\tilde{s}$ (see the insets in Fig.~\ref{FigFidSculpTelep}), corresponding to the following entangled resources: squeezed symmetric superposition of Fock states, squeezed Bell-like states, and two-mode squeezed states, optimized for the teleportation of input coherent states $|\beta\rangle$, and input Fock states $|1\rangle$. The plot styles are chosen as specified in Fig.~\ref{FigFidSculpTelep}.}
\label{FigvonNeumSTruncTwB}
\end{figure}

\begin{figure}[h]
\centering
\includegraphics*[width=17cm]{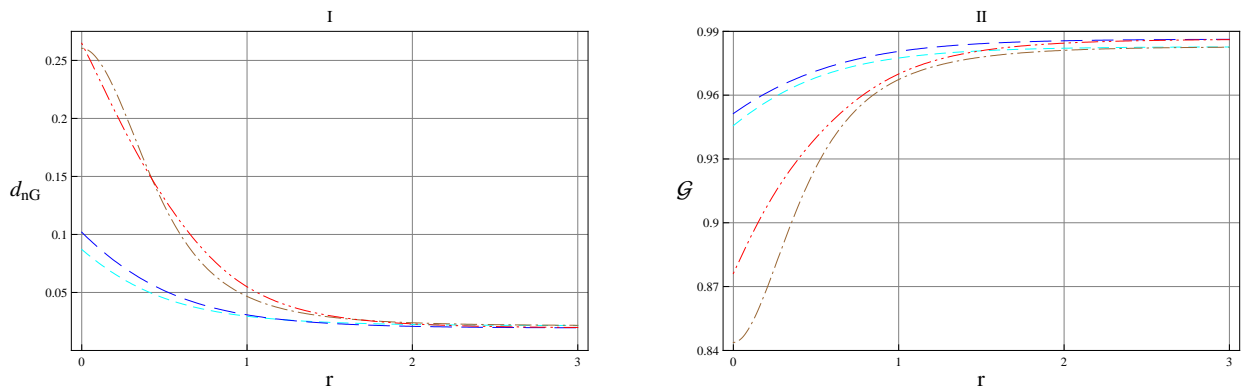}
\caption{Plot of the non-Gaussianity $d_{nG}$ (panel I)
and of the squeezed-vacuum affinity $\mathcal{G}$ (panel II)
as a function of $r$, with $s=\tilde{s}$
(see the insets in Fig.~\ref{FigFidSculpTelep}),
corresponding to the following entangled resources:
squeezed symmetric superposition of Fock states,
and squeezed Bell-like states,
optimized for the teleportation of input coherent states $|\beta\rangle$,
and input Fock states $|1\rangle$.
The plotstyles are chosen as specified in Fig.~\ref{FigFidSculpTelep}.
}
\label{FigNonGaussAffSTruncTwB}
\end{figure}

The standard Braunstein-Kimble CV teleportation protocol,
usually described in the domain of quantum optics
by the Wigner function formalism \cite{BraunsteinKimble},
can be expressed equivalently in different mathematical approaches \cite{TelepFormal1,TelepFormal2,TelepFormal3,MarianCVTelep} (
For a comprehensive review on CV quantum teleportation and information processing,
see Ref.~\cite{VanLoockBraunstein}).
Following a recent proposal, a very convenient description of the protocol can be given
in terms of the characteristic functions of the various
quantum states involved, including input, resource, and teleported states
\cite{MarianCVTelep}. Denoting by
$\rho_{in}$ and $\chi_{in}(\alpha_{in})$, respectively, the
single-mode input state to be teleported and the associated
characteristic function, and by $\rho_{12}$ and $\chi_{12}(\alpha_{1}\,,\alpha_{2})$,
respectively, the entangled two-mode resource
and its characteristic function, it can be shown that the characteristic function
$\chi_{out}(\alpha_{2})$ of the teleported state takes the
factorized form \cite{MarianCVTelep}:
\begin{equation}
\chi_{out}(\alpha_{2}) \,=\, \chi_{in}(\alpha_{2}) \;
\chi_{12}(\alpha_{2}^{*}\,,\alpha_{2}) \; .
\label{telepchiout}
\end{equation}
A commonly used measure of the success probability of a teleportation
protocol is the fidelity of teleportation, i.e. $\mathcal{F} \, = \,
Tr[\rho_{in}\rho_{out}]$. In the characteristic-function formalism,
the fidelity reads
\begin{equation}
\mathcal{F} \,=\,  \frac{1}{\pi} \int d^{2}\mathbf{\lambda} \;
\chi_{in}(\mathbf{\lambda}) \chi_{out}(-\mathbf{\lambda}) \,.
\label{Fidelitychi}
\end{equation}
In the following we will adopt Eq.~(\ref{Fidelitychi}) to analyze
the efficiency of the CV teleportation protocol. \\
In Ref.~\cite{CVTelepNoi}, the squeezed Bell-like states have been exploited
as non-Gaussian entangled resources optimizing the teleportation fidelity
for various sets of input states.
The general expression of squeezed Bell-like states reads:
\begin{equation}
|\psi\rangle_{SB} = [c_{1}^{2}+c_{2}^{2}]^{-1/2} S_{12}(\zeta)
\{c_{1} |0,0 \rangle_{12} + e^{i \theta} c_{2}
|1,1 \rangle_{12}\} ,
\label{squeezBell}
\end{equation}
where $S_{12}(\zeta) = e^{ -\zeta a_{1}^{\dag}a_{2}^{\dag} + \zeta
a_{1}a_{2}}$ is the two-mode squeezing operator, $\zeta=r e^{i\phi}$,
$|m_{1}\,,m_{2} \rangle_{12} \equiv |m_{1}\rangle_{1} \otimes |m_{2}\rangle_{2}$
is a two-mode Fock state, and the $c_{i}$s are real constants.
A convenient parametrization is $c_{1}=\cos\delta$ and $c_{2}=\sin\delta$.
At fixed squeezing parameter $r$, and at fixed phases $\phi=\pi$ and $\theta=0$, the entangled resource Eq.~(\ref{squeezBell}) is endowed with an additional phase
$\delta$, that can be freely adjusted and thus used to maximize the fidelity $\mathcal{F}_{SB}$:
\begin{equation}
\mathcal{F}_{opt}(r) \,=\, \max_{\delta} \;
\mathcal{F}_{SB}(r,\delta) \; .
\label{FidOptim}
\end{equation}
Such an optimization yields a remarkable enhancement in the success
probability of teleportation for various input states \cite{CVTelepNoi}.
The exact analytical expression $\mathcal{F}_{SB}(r,\delta)$ for the fidelity of teleportation of input coherent states using optimized squeezed Bell-like resources, i.e. Eq.~(\ref{FidoptSB}),
is reported in Appendix \ref{appendix}.
The optimal phase $\delta=\delta_{max}$, maximizing the fidelity $\mathcal{F}_{SB}(r,\delta)$ is given by:
\begin{equation}
\delta_{max} \,=\, \frac{1}{2} \arctan(1+e^{-2 r}) \; .
\label{deltamax}
\end{equation}
\\
It is now important to observe that the Bell-like contribution in Eq.~(\ref{squeezBell}) can be obtained
from the truncation at first order $(n=1)$ of twin-beam states,
i.e. $S_{12}(\xi)|0,0\rangle_{12} \propto
\sum_{n=0}^{\infty}(-e^{i\varphi}\tanh s)^{n} |n,n\rangle_{12}$,
with $\xi=e^{i\varphi}s$.
This fact can be more clearly understood if one chooses, in Eq.~(\ref{squeezBell}),
the parametrization $c_{1}=1$ and $c_{2}=\tanh s$ (with $\varphi=\theta+\pi$).
Thus, although being a non-Gaussian state, the squeezed Bell-like state
can be extracted by a Gaussian squeezed state by means of a truncation procedure;
in other words, it is a squeezed truncated-TwB.
However, the connection of entangled non-Gaussian resources with the Gaussian twin beams
may break down if we further generalize the squeezed Bell-like state (\ref{squeezBell})
to the following class of non-Gaussian states,
the two-mode squeezed symmetric superposition of Fock states:
\begin{equation}
|\psi\rangle_{SSF} =[c_{1}^{2}+c_{2}^{2}+c_{3}^{2}]^{-1/2} S_{12}(\zeta)\{c_{1} |0,0 \rangle_{12}
 + e^{i \theta_{2}} c_{2}
|1,1 \rangle_{12} + e^{i \theta_{3}} c_{3}
|2,2 \rangle_{12}\} ,
\label{squeezSSF}
\end{equation}
where $c_{i}$ and $\theta_{i}$ are real constants and phases, respectively.
With respect to state (\ref{squeezBell}), state (\ref{squeezSSF}) is not
constrained to belong to the class of squeezed truncated-TwB.
Namely, thanks to the additional four-photon term in the superposition, the class of squeezed symmetric superposition of Fock states includes the squeezed truncated-TwB as a special realization.
Now, it is very interesting to determine the subset of states $|\psi\rangle_{SSF}$
corresponding to an optimal fidelity of teleportation.
Obviously, it is expected that the exploitation of the resource (\ref{squeezSSF}) will also allow a further optimization of the teleportation protocol.
We compute the teleportation fidelity for an input coherent state $|\beta\rangle$ and an input single-photon Fock state $|1\rangle$.
A convenient parametrization of the $c_{i}$s in Eq.~(\ref{squeezSSF}) is provided by the hyperspherical coordinates in three dimensions: $c_{1}=\cos\delta_{1}$, $c_{2}=\sin\delta_{1} \cos\delta_{2}$, $c_{3}=\sin\delta_{1} \sin\delta_{2}$.
The two-mode characteristic function $\chi_{SSF}$ of the entangled resource (\ref{squeezSSF})
can be calculated as
\begin{equation}
\chi_{SSF}(\alpha_{1},\,\alpha_{2})=Tr[\rho D_{1}(\alpha_{1})D_{2}(\alpha_{2})]  =\,_{SSF}\langle\psi|D_{1}(\alpha_{1})D_{2}(\alpha_{2})|\psi\rangle_{SSF} \,,
\label{chiSSF}
\end{equation}
where $D_{i}(\alpha_{i})$ is the Glauber operator for the mode $i$.
The explicit expression for Eq.~(\ref{chiSSF}) can be easily written
by making use of the two-mode Bogoliubov transformations
\begin{equation}
S_{12}^{\dag}(\zeta)\, a_{i} \, S_{12}(\zeta)=\cosh r \, a_{i}
-e^{i\phi}\sinh r \, a_{j}^{\dag}, \, (i\neq j=1,2) \,,
\label{BogoliubovT}
\end{equation}
and of the relation
\begin{equation}
\langle m| D(\alpha) |n \rangle \,=\,
\left(\frac{n!}{m!}\right)^{1/2}\alpha^{m-n}e^{-\frac{1}{2}|\alpha|^{2}}
L_{n}^{(m-n)}(|\alpha|^{2}) \,,
\label{LaguerreFormula}
\end{equation}
where $L_{n}^{(m-n)}(\cdot)$ denotes the associated Laguerre polynomial.
The characteristic functions $\chi_{C}$ and $\chi_{F}$ corresponding to an input coherent state $|\beta\rangle$
and to an input single-photon Fock state $|1\rangle$, respectively, are given by:
\begin{eqnarray}
\chi_{C}(\alpha_{in}) \,&=&\,
e^{-\frac{1}{2}|\alpha_{in}|^{2}+2i Im[\alpha_{in}\beta^{*}]}
\; ,
\label{chiCohin}
\\
\chi_{F}(\alpha_{in}) & = & e^{-\frac{1}{2}|\alpha_{in}|^{2}} \,
(1-|\alpha_{in}|^{2}) \; .
\label{chiFockin}
\end{eqnarray}
Given Eqs.~(\ref{chiSSF}), (\ref{chiCohin}), and (\ref{chiFockin}), and using Eq.~(\ref{telepchiout}), the fidelities, defined by Eq.~(\ref{Fidelitychi}), for the two different input states can be analytically computed. The resulting expressions are functions of the parameters $r$, $\phi$, $\delta_{1}$, $\theta_{1}$, $\delta_{2}$, $\theta_{2}$, but we do not report them, as they are too long and cumbersome.
At fixed squeezing $r$, with maximally fixed phase $\phi=\pi$, the optimized fidelity is defined as
\begin{equation}
\mathcal{F}_{opt}=\max_{\mathcal{P}} \; \mathcal{F}(r,\pi,\delta_{1},\theta_{1},\delta_{2},\theta_{2})
\label{optimalFid}
\end{equation}
where $\mathcal{P}$ denotes the set of independent free parameters $\mathcal{P}=\{\delta_{1},\theta_{1},\delta_{2},\theta_{2}\}$.
Thus, once the class of single-mode input states is specified,
the optimization procedure of the available free parameters,
in order to maximize the fidelity, leads to a \emph{sculpturing
of the teleported output state}. We compute the optimal fidelities (\ref{optimalFid}), for the above two classes of input states, by numerical maximization of the analytical expressions of $\mathcal{F}$. First, we notice that the phases $\theta_{1}$ and $\theta_{2}$ can be set to zero as they do not contribute to an improvement of $\mathcal{F}$.
The procedure leads to the following remarkable result: Optimization over the set $\mathcal{P}$ yields maximal parameters compatible with \emph{a squeezed truncated-TwB as optimal entangled non-Gaussian resource}.
Thus, the optimal non-Gaussian resources for CV teleportation form a subset of the class
(\ref{squeezSSF}) of the form:
\begin{equation}
|\psi'\rangle_{SSF} = [1+\tanh^{2}s+\tanh^{4}s]^{-1/2} S_{12}(-r)\{|0,0 \rangle_{12}
+ \tanh s
|1,1 \rangle_{12} + \tanh^{2}s
|2,2 \rangle_{12}\} ,
\label{squeezSSF'}
\end{equation}
where, correspondingly to Eq.~(\ref{optimalFid}), the real parameter $s$ is fixed to the maximal value $s=\tilde{s}$, through the numerical maximization of the fidelity, i.e. $\mathcal{F}_{opt}=\max_{s}\mathcal{F}(r,s)$.
In Fig.~\ref{FigFidSculpTelep}, we show the behavior of $\mathcal{F}_{opt}$ as a function of $r$ for the input states $|\beta\rangle$ (Panel I) and $|1\rangle$ (Panel II), with a squeezed symmetric superposition of Fock states as entangled resource. For comparison, also the curves $\mathcal{F}_{opt}$, corresponding to the teleportation with a squeezed state and with a squeezed Bell-like state, are plotted.
We observe that, as expected, a further enhancement of the fidelity is obtained,
compared to the case discussed in Ref.~\cite{CVTelepNoi}. \\
In Ref.~\cite{CVTelepNoi}, in order to understand the properties of the optimized resources, three quantities have been investigated: the von Neumann entropy $E_{vN}$, in order to quantify the amount of entanglement present in the resource; the non-Gaussianity $d_{nG}$, in order to provide a measure of the non-Gaussian character of the resource; and the squeezed-vacuum affinity $\mathcal{G}$, in order to determine the degree of resemblance of the resource to a two-mode squeezed vacuum.
In Fig.~\ref{FigvonNeumSTruncTwB}, we plot $E_{vN}$ for the optimized non-Gaussian entangled resources Eqs.~(\ref{squeezBell}) and (\ref{squeezSSF'}), and for the twin beam.
It can be observed that all the curves exhibit very similar behaviors. At fixed $r$, and for a given input state, the optimized squeezed symmetric superposition of Fock states is the most entangled state, and the teleportation of a single-photon requires more entanglement than the teleportation of a coherent state.
However, we notice that the behavior of the von Neumann entropies is in agreement with the
behavior of the optimal fidelities; in fact, at fixed input state,
a non-Gaussian resource with a higher teleportation fidelity is associated with
a higher amount of entanglement content for any $r$..
Next, we examine the behavior of the measure of non-Gaussianity $d_{nG}$,
which is defined as the distance between the considered state $\rho$
and a reference Gaussian state $\rho_{G}$,
according to the following relation:
\begin{equation}
d_{nG} =
\frac{Tr[\rho^{2}]+Tr[\rho_{G}^{2}]-2Tr[\rho \rho_{G}]}{2
Tr[\rho^{2}]} ,
\label{nonGaussianity}
\end{equation}
where the Gaussian state $\rho_{G}$ is
completely determined by fixing for it the same covariance matrix
and the same first order statistical moments of the quadrature operators
associated to state $\rho$ \cite{GenoniNonGaussy}.
In Fig.~\ref{FigNonGaussAffSTruncTwB} panel I, we plot $d_{nG}$ for the entangled resources
Eqs.~(\ref{squeezBell}) and (\ref{squeezSSF'}), optimized for teleportation.
Looking at the behaviors of $d_{nG}$ corresponding to teleportation of the same input states,
we see that, both in the instance of input coherent states
and in the instance of input Fock states,
the curves cross at a certain value of $r$.
Therefore, according to the particular adopted measure of non-Gaussianity (\ref{nonGaussianity}),
a non-Gaussian resource with a higher teleportation fidelity does not exhibit, in general,
a higher non-Gaussian character for any $r$.
In order to quantify how a given state resembles a squeezed vacuum,
a measure of the \emph{squeezed-vacuum affinity} can be introduced \cite{CVTelepNoi}.
Such a quantity is defined as the following maximized overlap:
\begin{equation}
\mathcal{G} \,=\, \max_{\xi} |\,_{12}\langle -\xi|\psi_{res}(r)\rangle_{12}|^{2} \; ,
\label{OverlapTWB}
\end{equation}
where $|-\xi\rangle_{12}$ is a two-mode squeezed vacuum with real squeezing
parameter $-\xi$, and $|\psi_{res}(r)\rangle_{12}$ is any entangled two-mode
resource that depends uniquely on the squeezing $r$ as the only free parameter.
In Fig.~\ref{FigNonGaussAffSTruncTwB} panel II, we plot $\mathcal{G}$ for the entangled resources
Eqs.~(\ref{squeezBell}) and (\ref{squeezSSF'}), optimized for teleportation.
At fixed input states, the optimized squeezed symmetric superposition of Fock states
exhibits a greater squeezed-vacuum affinity for any $r$
with respect to the squeezed Bell-like state.
This result is in complete agreement with the behaviors of the corresponding
optimal teleportation fidelities and von Neumann entropies.
Therefore, comparing the behaviors associated with resources optimized
for the efficient teleportation of the same input states,
we observe that the following hierarchy can be established:
the enhancement of the teleportation fidelity corresponds to an enhancement of the entanglement
content, and to an enhancement of squeezed-vacuum affinity.
This hierarchy is not respected by the measure of non-Gaussianity $d_{nG}$.
This result shows that such a measure must be used and interpreted with care, case by case.
As expected, in the plots of Fig.~\ref{FigNonGaussAffSTruncTwB}
panel I and panel II, for sufficiently large $r$, the curves, associated with the same resource,
go to the same asymptotic saturation value. \\
The following conclusions can be drawn from the above discussion:
The squeezed truncated twin beam, with truncation at $n=2$ (four-photon term),
is an optimal entangled non-Gaussian resource suitable for sculptured quantum teleportation.
We argue that this result can be generalized to any order $n>2$ in expansions of the squeezing
operator at finite order. We thus conjecture that, in general, efficient entangled non-Gaussian
resources for CV quantum teleportation should be characterized by a suitable balance between the
entanglement content and the degree of affinity to squeezed vacuum states.

\section{CV teleportation with squeezed cat-like resources}
\label{SecSquezCatlike}

By replacing in Eq.~(\ref{squeezBell}) the two-photon term $|1,1\rangle_{12}$
with the two-mode symmetric coherent state
$|\gamma,\gamma \rangle_{12}$, with $\gamma$ complex amplitude
$(|\gamma,\gamma \rangle_{12}\equiv |\gamma\rangle_{1}\otimes |\gamma\rangle_{2})$,
the squeezed Bell-like state (\ref{squeezBell}) is converted in the following
squeezed cat-like superposition:
\begin{equation}
|\psi\rangle_{SC} = \mathcal{N}\, S_{12}(\zeta)
\{\cos\delta |0,0 \rangle_{12} + e^{i \theta} \sin\delta
|\gamma,\gamma \rangle_{12}\} ,
\label{squeezCat}
\end{equation}
where the normalization factor is $\mathcal{N}=\{1+ e^{-|\gamma|^{2}}\sin 2\delta \cos\theta\}^{-1/2}$.
We remark that the possibility of engineering states of the form $|\psi\rangle_{SC}$ is confirmed by the recent successful experimental realization of their single-mode equivalent \cite{GrangierCats}.
The characteristic function associated with Eq.~(\ref{squeezCat}), $\chi_{SC}(\alpha_{1},\alpha_{2})=\,_{SC}\langle\psi|D_{1}(\alpha_{1})D_{2}(\alpha_{2})|\psi\rangle_{SC},$ can be easily calculated, yielding a simple algebraic sum of Gaussian functions.
We can thus compute analytically the fidelity of teleportation using squeezed cat-like states (\ref{squeezCat}) as entangled resource, limiting the analysis, for simplicity, to the teleportation of input coherent states $|\beta\rangle$.
The maximization procedure is discussed in Appendix \ref{appendix},
including the analytical expression for the fidelity.
At fixed squeezing $r$, the optimal fidelity can be obtained
by maximizing on the real amplitude $|\gamma|$, as the only free parameter
\begin{equation}
\mathcal{F}_{opt}(r) \,=\, \max_{|\gamma|}\mathcal{F}_{SC}\left(r,|\gamma|\right) \,.
\end{equation}
\begin{figure}[ht]
\centering
\includegraphics*[width=8cm]{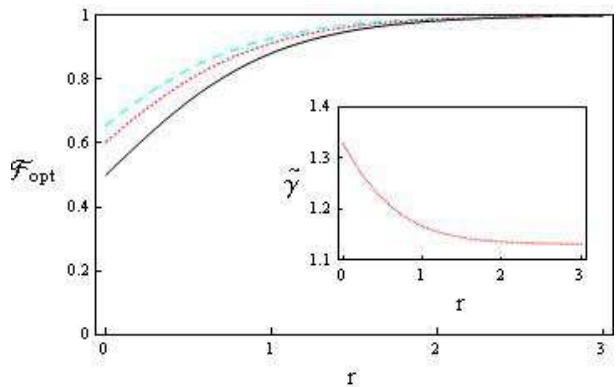}
\caption{Optimal fidelity of teleportation
$\mathcal{F}_{opt}$, as a function of the squeezing parameter
$r$, with $\phi=0$. The fidelity corresponds to the teleportation
of input coherent states $|\beta\rangle$ with a squeezed cat-like state
as entangled resource (dotted line).
For comparison, we plot the fidelities obtained with a squeezed vacuum
resource(full line) and with a squeezed Bell-like resource (dashed line).
The inset gives the maximal value of the parameter $|\gamma|=\tilde{\gamma}$
as a function of $r$.}
\label{FigFidSculpTelep2}
\end{figure}
In Fig.~\ref{FigFidSculpTelep2}, we show the behavior of $\mathcal{F}_{opt}$ as a function of $r$ for the input states $|\beta\rangle$, with an optimized squeezed cat-like state as entangled resource. For comparison, also the optimal fidelities, corresponding to the teleportation with Gaussian squeezed state
resources and with squeezed Bell-like states, are plotted.
Optimized squeezed cat-like non-Gaussian resources yield a significant improvement of the fidelity with respect to the two-mode squeezed vacuum, but are less efficient than squeezed Bell-like resources.
This result is in agreement with the behaviors of the von Neumann entropy, the non-Gaussianity,
and the squeezed-vacuum affinity,
reported in Figs.~\ref{FigvonNeumSqCat}, and \ref{FigdnGAffSqCat} panel I and panel II,
respectively.
In fact, the optimized squeezed cat-like superposition is less entangled,
less non-Gaussian (according to the particular measure $d_{nG}$),
and less affine to the squeezed-vacuum, than the optimized squeezed Bell-like state.
In particular, the optimized squeezed cat-like resource exhibits a marked reduction of $\mathcal{G}$,
if compared with the corresponding plots shown in Fig.~\ref{FigdnGAffSqCat} panel II;
in fact, $\mathcal{G}$ keeps lower than $71\%$, for any $r$.
\begin{figure}[t]
\centering
\includegraphics*[width=8cm]{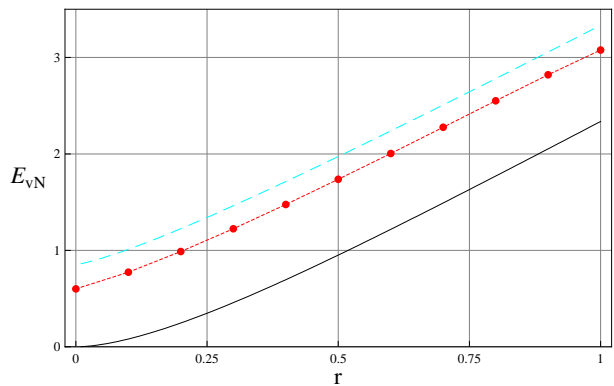}
\caption{Behavior of the von Neumann entropy $E_{vN}$ as a function of $r$, with $|\gamma|=\tilde{\gamma}$ (see the inset in Fig.~\ref{FigFidSculpTelep2}) for the optimized squeezed cat-like state (dotted line).
For comparison, we also plot the quantities $E_{vN}$
corresponding to a squeezed vacuum (full line),
and to a squeezed Bell-like state (dashed line).}
\label{FigvonNeumSqCat}
\end{figure}
\begin{figure}[t]
\centering
\includegraphics*[width=17cm]{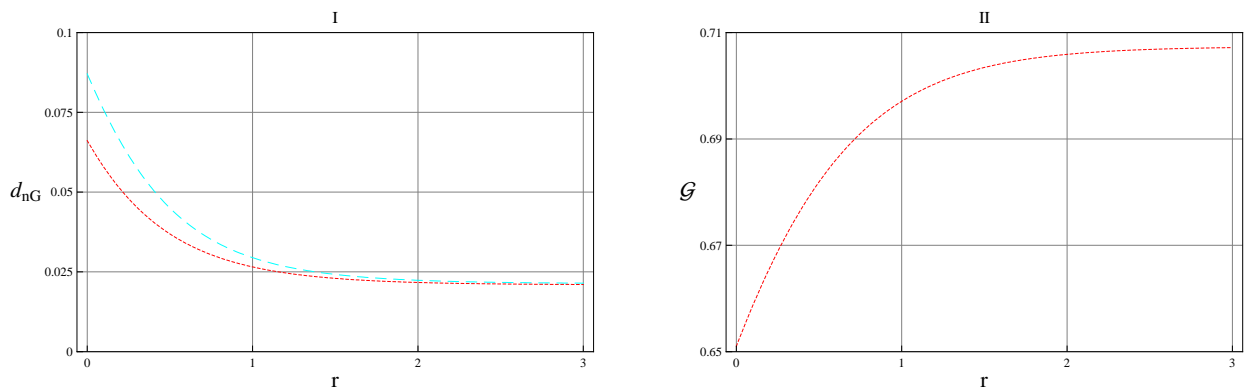}
\caption{Behavior of the non-Gaussianity $d_{nG}$ (panel I) and of
the squeezed-vacuum affinity $\mathcal{G}$ (panel II) as a function of $r$,
with $|\gamma|=\tilde{\gamma}$ (see the inset in Fig.~\ref{FigFidSculpTelep2})
for the optimized squeezed cat-like state (dotted line).
For comparison, in panel I we plot $d_{nG}$ for the optimized squeezed Bell-like state (dashed line).}
\label{FigdnGAffSqCat}
\end{figure}

\section{CV teleportation with noisy non-Gaussian states}
\label{SecNonGaussMixed}

\begin{figure}[h]
\centering
\includegraphics*[width=17cm]{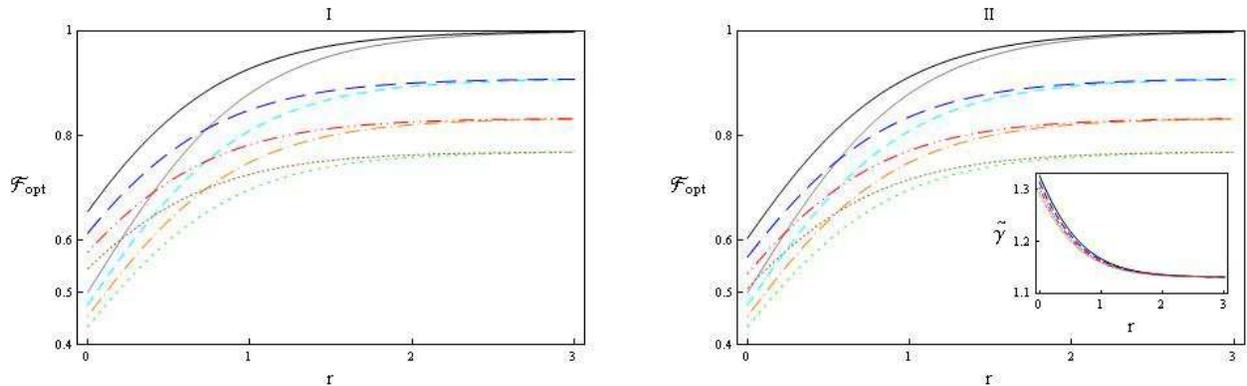}
\caption{Optimal fidelity $\mathcal{F}_{opt}$, associated with the teleportation
of input coherent states $|\beta\rangle$ with mixed squeezed Bell-like resources (panel I)
and mixed squeezed cat-like resources (panel II),
as a function of the squeezing $r$, for several choice of the thermal parameters
$n_{th,1}=n_{th,2}=n_{th}$.
The curves representing the fidelities associated with mixed non-Gaussian
entangled resources are plotted with the following plot style:
$n_{th}=0$ (full black line), $n_{th}=0.05$ (long-dashed line),
$n_{th}=0.10$ (double-dot dashed line), and $n_{th}=0.15$ (dotted line).
For comparison, in both panels, we also plot the fidelities associated to the mixed squeezed vacuum
entangled resources, with $n_{th}=0$ (full gray line), $n_{th}=0.05$ (dashed line),
$n_{th}=0.10$ (dot-dashed line), and $n_{th}=0.15$ (long-dotted line).
The inset in panel II gives the maximal value of the parameter $|\gamma|=\tilde{\gamma}$
as a function of $r$, with $n_{th}=0,\,0.05,\,0.1,\,0.15$
(same plot style as for the fidelities).}
\label{FigFidTelepMixedNG}
\end{figure}

\begin{figure}[h]
\centering
\includegraphics*[width=8cm]{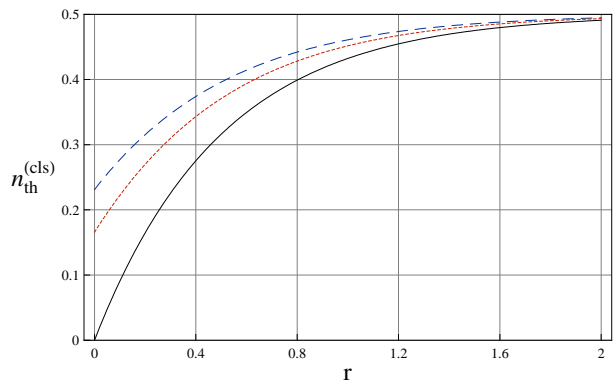}
\caption{Behavior of the threshold value $n_{th}^{(cls)}$ as a function of $r$,
for mixed squeezed Bell-like resources (dashed line),
mixed squeezed cat-like resources (dotted line), and
mixed squeezed vacuum (full line).}
\label{FignthClassicalFid}
\end{figure}

In this Section, we extend the procedures described above to the
instance in which, while keeping the protocol ideal, the entangled resources
are affected by thermal noise.
In order to describe how a mixed resource arises from a pure one when a thermal background
is turned on, we consider fields corresponding to independent contributions and converging
on a common spatial volume. We follow a very effective scheme proposed by Glauber, based on
the $P$-function formalism \cite{Glauber}.
The superposition of two independently generated fields, both admitting a $P$
representation, is simply the convolution of the individual $P$-functions:
\begin{equation}
P(\xi)=\int\int d^{2}\xi' P_{1}(\xi-\xi')\,P_{2}(\xi') \, .
\end{equation}
As demonstrated in Ref.~\cite{Rockower} (see also Ref.~\cite{Marianthermal}),
if \emph{at least one} of the independent fields admits a valid $P$ representation,
then the normally ordered quantum characteristic
function $\chi_{N}(\alpha)$ of the superposition is given by
the multiplication of the individual characteristic functions:
\begin{equation}
\chi_{N}(\alpha) \,=\, \chi_{N}^{(1)}(\alpha) \,\chi_{N}^{(2)}(\alpha) \; .
\label{charfuncmultipl}
\end{equation}
Thus, the relation (\ref{charfuncmultipl}) can be applied in the case of the
superposition of a generic field $\rho$ with a single-mode thermal field
\begin{equation}
\rho_{th}^{(1)} = \sum_{k}\frac{n_{th}^{k}}{(1+n_{th})^{k+1}} |k \rangle \langle k| \; ,
\label{thermalfield}
\end{equation}
whose
normally ordered characteristic function is a Gaussian function of the form
\begin{equation}
\chi_{N}^{(th)}(\alpha) \,=\, e^{-n_{th}|\alpha|^{2}} \; ,
\label{chinormthermal}
\end{equation}
where $n_{th}$ denotes the average number of thermal photons.
Generalizing the formula (\ref{charfuncmultipl}) to the two-mode instance,
we consider the superposition of a two-mode non-Gaussian
state $\rho=|\psi\rangle_{nG}\,_{nG}\langle\psi|$ and a two-mode thermal state
$\rho_{th}^{(2)} = \rho_{th1}^{(1)} \otimes \rho_{th2}^{(1)}$.
Next, we specialize the analysis to the two-mode squeezed Bell-like state
and to the two-mode squeezed cat-like state, given by Eqs.~(\ref{squeezBell}) and (\ref{squeezCat}),
respectively.
The resulting field is then described by the symmetrically ordered characteristic function
\begin{equation}
\chi_{nG}^{(th)}(\alpha_{1},\alpha_{2}) \,=\,  e^{-n_{th,1}|\alpha_{1}|^{2}-n_{th,2}|\alpha_{2}|^{2}}
\, \chi_{nG}(\alpha_{1},\alpha_{2}) \; ,
\label{chinGth}
\end{equation}
where $n_{th,i}$ is the thermal parameter associated with mode $i$ $(i=1,2)$,
and $\chi_{nG}(\alpha_{1},\alpha_{2})$ denotes the characteristic function associated with the
considered non-Gaussian state.
The state described by Eq.~(\ref{chinGth}) is mixed as soon as $n_{th, i} \neq 0$.
Of course, for $\delta=0$ we obtain the characteristic function for the
superposition of a two-mode squeezed vacuum state
$\rho=|\zeta\rangle\,\langle\zeta|$ and a two-mode thermal state $\rho_{th}^{(2)}$.
As in the previous Sections, we can now exploit the mixed non-Gaussian resource, described by the
characteristic function (\ref{chinGth}), and use Eq.~(\ref{chiCohin}) to
calculate the fidelity of teleportation, for instance in the case of input coherent states.
The explicit analytical expressions for the fidelities of teleportation of input coherent states
either with mixed squeezed Bell-like resources or with mixed squeezed cat-like resources,
i.e. $\mathcal{F}_{SB}^{(th)}(r,n_{th,1},n_{th,2},\delta)$ and $\mathcal{F}_{SC}^{(th)}(r,n_{th,1},n_{th,2},|\gamma|)$,
are reported in Appendix \ref{appendix}.
In the case of mixed squeezed Bell-like states,
at given thermal numbers $n_{th,1}$, $n_{th,2}$, and fixed squeezing parameter $r$,
the optimal fidelity is defined as:
\begin{equation}
\mathcal{F}_{opt}(r,n_{th,1},n_{th,2}) \,=\, \max_{\delta} \;
\mathcal{F}_{SB}^{(th)}(r,n_{th,1},n_{th,2},\delta) \; ,
\label{FidthSBOptim}
\end{equation}
and the optimal angle $\delta=\delta_{max}$ is easily computed:
\begin{equation}
\delta_{max} \,=\, \frac{1}{2} \arctan\Big(1+\frac{e^{-2r}}{1+n_{th,1}+n_{th,2}}\Big) \,.
\label{deltamaxmixed}
\end{equation}
In the case of mixed squeezed cat-like states,
at given thermal numbers $n_{th,1}$, $n_{th,2}$, and fixed squeezing parameter $r$,
the optimal fidelity is defined as:
\begin{equation}
\mathcal{F}_{opt}(r,n_{th,1},n_{th,2}) \,=\, \max_{|\gamma|} \;
\mathcal{F}_{SC}^{(th)}(r,n_{th,1},n_{th,2},|\gamma|) \; .
\label{FidthSCOptim}
\end{equation}
In Eq.~(\ref{FidthSCOptim}) the maximization can be performed numerically.
In Fig.~\ref{FigFidTelepMixedNG}, we plot the optimal fidelities for the mixed non-Gaussian resources
as a function of squeezing for various choices of the thermal
parameters $n_{th,1}=n_{th,2}=n_{th}$.
We observe that, as expected, the fidelity decreases for increasing $n_{th}$:
the thermal noise sensibly reduces the success probability of teleportation.
However, at fixed squeezing and thermal parameters, the non-Gaussian resources
always perform better than the Gaussian twin beam. Moreover, also in the mixed instance,
mixed squeezed Bell-like states perform better than mixed squeezed cat-like states.
Importantly, for the chosen realistic values of $n_{th}$, the
fidelity associated with the mixed non-Gaussian resources never drops below
the threshold of classical teleportation with maximal fidelity
$\mathcal{F}_{cls}^{max} = 0.5$. On the other
hand, one can look for the threshold value $n_{th}^{(cls)}(r)$ such that, at fixed $r$,
one has $\mathcal{F}_{opt} = \mathcal{F}_{cls}^{max} =\frac{1}{2}$.
In Fig.~\ref{FignthClassicalFid} we plot $n_{th}^{(cls)}$ as a function of $r$.
We see that the threshold value $n_{th}^{(cls)}$ associated to non-Gaussian resources
is larger than that associated to Gaussian ones, and in any case sensibly larger than
the reasonable realistic values we have considered.
On the contrary, the fidelity associated with the mixed squeezed vacuum state,
at low values of $r$, falls below the classical threshold, unless optimization
is performed on the local degrees of squeezing, without modifying the Braunstein-Kimble
protocol \cite{GaussianOptimization}.

\section{Conclusions}
\label{Conclusions}

We have analyzed the performance of the Braunstein-Kimble continuous-variable
quantum teleportation protocol using two particular classes of non-Gaussian resources,
i.e. the squeezed truncated twin beam and the squeezed cat-like states.
The extra freely tunable parameters admitted by these resources in comparison
to standard Gaussian resources allow for an optimization procedure in order
to maximize the fidelity of teleportation.
For a given set of input states, such a procedure yields a
sculptured output teleported state.
In order to interpret the results and explain the enhancement of the fidelity,
we have characterized and compared
resources according to their degrees of entanglement, non-Gaussianity,
and squeezed-vacuum affinity.
We have shown that the optimized non-Gaussian resources possess
both a high entanglement content and a high degree of squeezed-vacuum affinity.
Finally, we have studied the efficiency of CV teleportation of input coherent states using,
as entangled resources, mixed non-Gaussian states of the radiation field.
We have shown that, although the thermal noise strongly affects the success probability of teleportation,
the resources provided by the optimized mixed-state extension of pure non-Gaussian resources
guarantee sufficiently high fidelities, for realistic values of the average thermal photon number
$n_{th}$ and of the squeezing $r$. In particular, they assure better performances with respect to the
mixed extensions of pure Gaussian twin-beam states. \\
The present analysis can be generalized by considering more general
non-Gaussian resources and fully realistic situations that involve simultaneously nonideal Bell
measurements and decoherence induced by propagation in noisy channels.
A further goal will be the study of optimal teleportation with non-Gaussian resources
of two-mode and multimode states \cite{twomodeinputTelep}, and the optimization of
teleportation protocols with non-Gaussian resources with respect to {\em local} parameters
and properties, extending the existing schemes for the local optimization of Gaussian
resources, that yield a one-to-one correspondence between the content of entanglement
present in a resource and maximal nonclassical output fidelity \cite{GaussianOptimization}.

\appendix

\section{Analytical expressions for the fidelities}
\label{appendix}

Here we report the analytical expressions of the fidelities of teleportation of input
coherent states with the pure and the mixed non-Gaussian squeezed Bell-like states as
resources (we do not report the whole calculation that is straightforward but somewhat cumbersome).
The fidelity $\mathcal{F}_{SB}(r,\delta)$ for the squeezed Bell-like state (\ref{squeezBell}) reads:
\begin{equation}
\mathcal{F}_{SB}(r,\delta) \,=\, \frac{1+e^{2r}+e^{4r}+e^{2r}\cos
2 \delta+(1+e^{2r})\sin 2\delta}{e^{-2r}(1+e^{2r})^{3}} \,.
\label{FidoptSB}
\end{equation}
For $\delta=0$, the state $|\psi\rangle_{SB}$ reduces to
the twin beam, and Eq.~(\ref{FidoptSB}) gives the well known formula:
\begin{equation}
\mathcal{F}_{TwB}(r)\equiv\mathcal{F}_{SB}(r,0) \,=\, \frac{1}{1+e^{-2r}} \,.
\end{equation}
The fidelity $\mathcal{F}_{SB}^{(th)}(r,n_{th,1},n_{th,2},\delta)$ for the mixed
squeezed Bell-like state reads:
\begin{equation}
\mathcal{F}_{SB}^{(th)}(r,n_{th,1},n_{th,2},\delta) \,=\,
\frac{1+e^{2r}f_{th}+e^{4r}f_{th}^{2}+e^{2r}f_{th}\cos 2\delta+[1+e^{2r}f_{th}]\sin
2\delta}{e^{-2r}[1+e^{2r}f_{th}]^{3}} \,,
\label{FidSBMixed}
\end{equation}
where we have defined
\begin{equation}
f_{th} \,=\, 1+n_{th,1}+n_{th,2} \,.
\label{fth}
\end{equation}
Of course, it is easy to check that, in the absence of thermal fields, i.e. $n_{th,1}=n_{th,2}=0$,
we get $f_{th}=1$, and the fidelity $\mathcal{F}_{SB}^{(th)}$ reduces to $\mathcal{F}_{SB}$.
Notice that, in the limit of large thermal parameter, the fidelity of teleportation (\ref{FidSBMixed})
is strongly suppressed and vanishes asymptotically.

Next we give the analytical expressions of the fidelities of teleportation of input
coherent states with pure and mixed non-Gaussian squeezed cat-like states as
resources.
By putting, as usual, the phases $\theta$ and $\phi$ to zero,
the fidelity for the squeezed cat-like state (\ref{squeezCat}) reads:
\begin{equation}
\mathcal{F}_{SC}'(r,\delta,\gamma) \,=\, \frac{\cos^{2}\delta +
e^{\frac{(\gamma-\gamma^{*})^{2}}{1+e^{2r}}}\sin^{2}\delta+
e^{-|\gamma|^{2}}\left(e^{\frac{\gamma^{2}}{1+e^{2r}}}+e^{\frac{\gamma^{*2}}{1+e^{2r}}}\right)
\sin\delta\cos\delta}{(1+e^{-2r})(1+e^{-|\gamma|^{2}}\sin 2\delta)} \,.
\label{FidelitySqueezKat}
\end{equation}
It is worth noticing that, for $\delta\rightarrow 0$ and/or $\gamma\rightarrow 0$, the fidelity (\ref{FidelitySqueezKat}) reduces to the well known expression $\mathcal{F}_{TwB}(r)$,
holding for a twin beam resource.
A preliminary numerical optimization procedure for Eq.~(\ref{FidelitySqueezKat})
allows to fix the parameters $\arg\gamma$ and $\delta$ to the values $\arg\gamma = 0$ and $\delta=\frac{\pi}{4}$, leading to a simplification of the fidelity to the formula:
\begin{equation}
\mathcal{F}_{SC}\left(r,|\gamma| \right) \,=\, \frac{1+
e^{\frac{-|\gamma|^{2}}{1+e^{-2r}}}}{(1+e^{-2r})(1+e^{-|\gamma|^{2}})} \,.
\label{FidelitySqueezKat2}
\end{equation}
Thus, at fixed squeezing $r$, the maximization can be carried out on the real amplitude $|\gamma|$, as the only free parameter, i.e. $\mathcal{F}_{opt}=\max_{|\gamma|}\mathcal{F}_{SC}\left(r,|\gamma|\right)$.
In the particular instance of zero squeezing $r=0$, it is easy to find value of the optimal fidelity $\mathcal{F}_{opt}=[4(\sqrt{2}-1)]^{-1}\simeq 0.6035$, for $|\gamma|=\ln^{1/2}(\sqrt{2}-1)^{-2}\simeq 1.3276$.
Considering the mixed instance, at fixed angle $\delta=\frac{\pi}{4}$
and phases $\phi=\pi$, $\theta=0$, $\arg\gamma =0$, the fidelity $\mathcal{F}_{SC}^{(th)}(r,n_{th,1},n_{th,2},|\gamma|)$ for the mixed
squeezed cat-like state reads:
\begin{equation}
\mathcal{F}_{SC}^{(th)}(r,n_{th,1},n_{th,2},|\gamma|) \,=\,
\frac{1 +
e^{-|\gamma|^{2}}e^{\frac{|\gamma|^{2}}{(1 + e^{2r} f_{th})}}}{e^{-2r}(1 + e^{2r} f_{th})(1+e^{-|\gamma|^{2}})} \,,
\label{FidelitySCMixed}
\end{equation}
where $f_{th}$ is defined in Eq.~(\ref{fth}).

\end{document}